\documentclass[12pt]{article}
\begin{document}
\thispagestyle{empty}
\begin{center} {\Large\bf Aging of the Universe and the fine
structure constant}

\bigskip\bigskip
\centerline{M. Kozlowski$^{\rm a}$, J. Marciak-Kozlowska$^{\rm
b,c}$}

\bigskip
\bigskip
\bigskip
\bigskip
\noindent{\llap{$^{\rm a}$~} Science Teachers College and
Institute of Experimental Physics, Warsaw University,
   Ho\.{z}a~69, 00-681 Warsaw, Poland, e-mail:mirkoz@fuw.edu.pl}

\noindent{\llap{$^{\rm b}$~}Institute of Electron Technology,
   Al.~Lotnik\'{o}w~32/46, 02-668 Warsaw, Poland}
   \end{center}

\hbox to 5cm{\hsize=5cm\vbox{\ \hrule}}\par \noindent{\llap{$^{\rm
c}$~}Author to whom correspondence should be addressed.}

\begin{abstract}
In this paper the aging of the Universe is investigated in the
frame of quantum hyperbolic heat transport equation. For the open
universe, when $t\to \infty$, $\hbar\to\infty$, $c\to 0$ and fine
structure constant, $\alpha$, is constant.

Key words: Quantum heat transport; Open universe; Fine structure
constant.
\begin{description}
\item[PACS98.62.Py]{Distances, redshifts, radial velocity, spatial
distribution of galaxies}
\item[PACS98.80.Bp]{Origin and formations of
the Universe}
\item[PACS98.80Cq]{Particle-theory and field theory
models of the early Universe (including cosmic pancakes, chaotic
phenomena, inflationary universe, etc.}
\end{description}
\end{abstract}

\newpage
\section{Introduction}
The distant future of the Universe is dramatically different
depending on whether it expands forever, or it stops expanding at
some future time and recolapses. The long term future of life and
civilization has been discussed by J.~N.~Islam~\cite{1} and
F.~Dyson~\cite{2}. In this paper we will study the aging of the
open universe in which $t\to\infty$. Starting from the quantum
hyperbolic heat transfer equation we argue that the Planck time
$\tau_P=\left(\frac{\hbar G}{c^5}\right)^{1/2}$ is the border
between the time reversible universe $t<\tau_P$ and universe with
time arrow for $t>\tau_P$. For time $t\to\infty$ the prevaling
thermal process for thermal phenomena in the universe is the
diffusion with diffusion constant $D_P=\left(\frac{\hbar
G}{c}\right)^{1/2}$, i.e. for $t\to\infty$, $D_P\to\infty$. From
formula for $D_P$ we conclude that for $t\to\infty$,
$\hbar\to\infty$ and $c\to 0$. In that case from formula for fine
structure constant~$\alpha$ we obtain $\alpha=$ constant for
$t\to\infty$. This result does not exclude the observed very small
change of $\alpha$ for redshift $0.5<z<3.5$~\cite{5, 6}. The
theory with variable $c$ was considered by J.~Magueijo~\cite{8}.
It seems quite interesting that in our scenario, $t\to\infty$,
$c\to 0$, $\hbar\to\infty$ the aging universe will be more and
more quantum Universe.
\section{The time arrow in a Planck gas}
The enigma of Planck era i.e., the event characterized by the
Planck time, Planck radius and Planck mass, is very attractive for
speculations. In this paper, we discuss the new interpretation of
Planck time. We define Planck gas -- a gas of massive particles
all with masses equal the Planck mass $M_P=\left(\frac{\hbar
c}{G}\right)^{1/2}$ and relaxation time for transport process
equals the Planck time $\tau_P=\left(\frac{\hbar
G}{c^5}\right)^{1/2}$. To the description of a thermal transport
process in a Planck gas, we apply the quantum Heaviside heat
transport equation~(QHH)~\cite{3}
    \begin{equation}
    \frac{\lambda_B}{v_h}\frac{\partial^2T}{\partial
    t^2}+\frac{\lambda_B}{\lambda}\frac{\partial T}{\partial
    t}=\frac{\hbar}{M_P}\nabla^2T.\label{eq1}
    \end{equation}
In Eq.~({\ref{eq1}) $M_P$ is the Planck mass, $\lambda_B$ the de
Broglie wavelength and $\lambda$ mean free path for Planck mass.
The Eq.~(\ref{eq1}) describes the dissipation of the thermal
energy induced by a temperature gradient $\nabla T$. Recently, the
dissipation of the thermal energy in the cosmological context
(e.g. viscosity) was described in the frame of EIT (Extended
Irreversible Thermodynamics)~\cite{2}. With the simple choice for
viscous pressure, it is shown that dissipative signals propagate
with the light velocity~$c$. Considering that the relaxation time
$\tau$ is defined as~\cite{3}
    \begin{equation}
    \tau=\frac{\hbar}{M_Pv_h^2},\label{eq2}
    \end{equation}
for thermal wave velocity $v_h=c$ one obtains
    \begin{equation}
    \tau=\frac{\hbar}{M_P c^2}=\left(\frac{\hbar
    G}{c^5}\right)^{\frac12}=\tau_P,\label{eq3}
    \end{equation}
i.e. \textit{the relaxation time is equal to the Planck time}
$\tau_P$. The gas of massive particles with masses equal to the
Planck mass $M_P$, and relaxation time $\tau_P$ we will define as
the Planck gas.

According to the results of paper~\cite{3} we define the quantum
of the thermal energy, \textit{heaton} for the Planck gas as
    \begin{equation}
    E_h=\hbar\omega_P=\frac{\hbar}{\tau_P}=\left(\frac{\hbar
    c}{G}\right)^{\frac12}c^2=M_Pc^2,\label{eq4}
    \end{equation}
i.e.
    \begin{equation}
    E_h=\hbar\omega_P=10^{19} {\rm GeV}.\label{eq5}
    \end{equation}
With formula~(\ref{eq2}) and $v_h=c$ we calculate the mean free
path~$\lambda$, viz
    \begin{equation}
    \lambda=v_h\tau_P=c\tau_P=\left(\frac{\hbar
    G}{c^3}\right)^{\frac12}.\label{eq6}
    \end{equation}
From formula~(\ref{eq6}) we conclude that mean free path for a
Planck gas is equal to the Planck radius. For a Planck mass, we
can calculate the de Broglie wavelength
    \begin{equation}
    \lambda_B=\frac{\hbar}{M_P
    v_h}=\left(\frac{G\hbar}{c^3}\right)^{\frac12}=\lambda.\label{eq7}
    \end{equation}
As it is defined in paper~\cite{3} Eq.~(\ref{eq7}) describes the
quantum limit of heat transport. When formulae~(\ref{eq6}) and
(\ref{eq7}) are substituted in Eq.~(\ref{eq1}) we obtain
    \begin{equation}
    \tau_P\frac{\partial^2 T}{\partial t^2}+\frac{\partial
    T}{\partial t}=\frac{\hbar}{M_P}\nabla^2 T.\label{eq8}
    \end{equation}
Equation~(\ref{eq8}) is the quantum hyperbolic heat transport
equation for a Planck gas. It can be written as
    \begin{equation}
    \frac{\partial^2 T}{\partial t^2}+\left(\frac{c^5}{\hbar
    G}\right)^{\frac12}\frac{\partial T}{\partial
    t}=c^2\nabla^2T\label{eq9}
    \end{equation}
The quantum hyperbolic heat equation~(\ref{eq9}) as a hyperbolic
equation sheds light on the time arrow in a Planck gas. When QHT
is written in the equivalent form
    \begin{equation}
    \tau_P\frac{\partial^2 T}{\partial t^2}+\frac{\partial
    T}{\partial t}=D_P\nabla ^2 T,\label{eq10}
    \end{equation}
where $D_P=\left(\frac{\hbar G}{c}\right)^{\frac12}$ is the
diffusion coefficient for a Planck gas, then for time period
shorter then $\tau_P$ we have preserved time reversal for thermal
processes, viz.
    \begin{equation}
    \frac{1}{c^2}\frac{\partial^2 T}{\partial t^2}=\nabla^2
    T.\label{eq11}
    \end{equation}
For the aging of the universe i.e for $t\to\infty$, $t\gg\tau_P$
the time reversal symmetry is broken
    \begin{equation}
    \frac{\partial T}{\partial t}=\left(\frac{\hbar
    G}{c}\right)^{\frac12}\nabla^2 T.\label{eq12}
    \end{equation}
These new properties of Eq.~(\ref{eq9}) open up new possibilities
for the interpretation of the Planck time. Before $\tau_P$,
thermal processes in Planck gas are symmetrical in time. After
$\tau_P$, i.e. for $t\to\infty$ the time symmetry is broken.
Moreover gravitation is activated after $\tau_P$ and this fact
creates an arrow of time~(\ref{eq12}).

It is well known that the equation~(\ref{eq11}) is invariant under
Lorentz group transformation  whereas equation~(\ref{eq12}) is
not. The time border between two processes domination waves and
diffusion is the $\tau_P$. On the other hand $\tau_P$ as the time
period is dependent on the observer velocity, i.e. can in
principle be different for different observers. The wayout which
solves the contradictory is to assume that $\tau_P$ is invariant
under Lorentz transformation. Considering that Planck length is
equal
    \begin{equation}
    L_P=c\tau_P,\label{eq13}
    \end{equation}
we obtain that $L_P$, Planck length is invariant under Lorentz
transformation. This conclusion is in harmony with the results of
G.~Amelino-Camelia paper~\cite{4}.
\section{Inconstancy of the fine structure constant,~$\alpha$?}
The contemporary observational method can compare the value of
fine structure constant $\alpha=\frac{1}{137}$ in different ages
of the universe~\cite{5, 6}. For the sources lying between
redshifts 0.5 and 3.5 as a whole, the observed shifts is
    \begin{equation}
    \frac{\Delta \alpha}{\alpha}=\frac{[\alpha(z)-\alpha({\rm now})]}{\alpha({\rm now})}
    =(-0.72\pm0.18)10^{-5}.\label{eq14}
    \end{equation}
If one converts this into a rate of charge of $\alpha$ with time
it amounts to about
    \begin{equation}
   \frac{(\textrm{rate of change of}\, \alpha)}{(\textrm{current
   value of}\, \alpha)}=5\cdot 10^{-16} \textrm{per year}\label{eq15}
   \end{equation}
One of the author of the papers~\cite{5, 6} once write on the
constant of nature~\cite{7}:

\begin{center}
    \parbox{10cm}{\textit{There is something attractive about permanance. We feel
    instinctively that things that have remained unchanged for
    centuries must posses some attribute that is instrinsically
    good\ldots. And despite the constant flux of changing events,
    we feel that the world possesses some invariant bedrock where
    general aspect the same}.}
    \end{center}

I still share this point of view and assume $\alpha$ is constant
through the evolution of universe. But the aging of the Universe
means the transition from wavy motion to the diffusion i.e the
diffusion is dominant for $t\to\infty$. The growing influence of
diffusion term in equation~(\ref{eq2}) means:
    $$
    D_P=\left(\frac{\hbar G}{c}\right)^{\frac12}\to \infty, \qquad {\rm for}
    \qquad t\to\infty
    $$
i.e. (when $G=$ constant) $\hbar\to\infty$, $c\to 0$ for
$t\to\infty$. In this scenario $\alpha=\frac{e^2}{\hbar c}$ can be
constant through the life of the universe -- our Universe.

One of the conclusion that for $t\to\infty$, $c\to 0$ is in
harmony with new results of Jo\~{a}o Magueijo~\cite{8} on the
varying of the light sped. It is interesting to observe that in
our scenario, i.e. $c\to 0$, $\hbar\to\infty$ for $t\to\infty$ the
aging Universe will be more and more quantum Universe with
prevaling quantum effects over the classical behaviour. Who knows?


\newpage
\begin{thebibliography}{99}
\bibitem{1}J.~N.~Islam, \textit{An Introduction to Mathematical
Cosmology}, CUP, 2002.
\bibitem{2}F.~J.~Dyson, \textit{Rev.~Mod.~Phys.}, \textbf{51},
(1979), p.~447.
\bibitem{3}M. Kozlowski, J. Marciak-Kozlowska,
\textit{From Quarks to Bulk Matter},  Hadronic Press, 2001.
\bibitem{4}G. Amelino-Camelia, \textit{Physics Letters~B},
\textbf{510}, (2001), p.~255.
\bibitem{5}J.~K.~Webb et al., \textit{ Phys. Rev. Lett.},
\textbf{82}, (1999), p.~884.
\bibitem{6}J.~K.~Webb et al., \textit{ Phys. Rev. Lett.},
\textbf{87}, (2001), p.~091301.
\bibitem{7}J.~D.~Barrow, \textit{Theories of Everything}, Fawcett
Columbine, N.Y. 1991, p.~117.
\bibitem{8}Jo\~{a}o Magueijo, \textit{Faster than the Speed of
Light}, Perseus Publishing, 2003.
\end{thebibliography}
\end{document}